\documentclass[fleqn,usenatbib,letter]{mnras}

\usepackage{newtxtext,newtxmath}
\usepackage[T1]{fontenc}

\DeclareRobustCommand{\VAN}[3]{#2}
\let\VANthebibliography\thebibliography
\def\thebibliography{\DeclareRobustCommand{\VAN}[3]{##3}\VANthebibliography}

\usepackage{graphicx}	
\usepackage{amsmath}	
\usepackage{amssymb}	
\usepackage{wasysym}           
\usepackage{graphicx}
\usepackage{epstopdf}
\usepackage{mathrsfs}
\usepackage{anyfontsize}
\usepackage{natbib}
\usepackage{color}
\usepackage{lipsum}
\usepackage{diagbox}

\title[Nearly isothermal shocks]{The structure of nearly isothermal, adiabatic shockwaves}

\author[E.~R.~Coughlin]{
Eric R.~Coughlin$^{1}$\thanks{E-mail: eric.r.coughlin@gmail.com} \\ 
$^{1}$Department of Astrophysical Sciences, Princeton University, Princeton, NJ 08544, USA
}

\date{Accepted XXX. Received YYY; in original form ZZZ}

\pubyear{2020}

\begin{document}
\label{firstpage}
\pagerange{\pageref{firstpage}--\pageref{lastpage}}
\maketitle

\begin{abstract}
An explosively generated shockwave with time-dependent radius $R(t)$ is characterized by a phase in which the shocked gas becomes radiative with an effective adiabatic index $\gamma \simeq 1$. Using the result that the post-shock gas is compressed into a shell of width $\Delta R/R \simeq  \delta$, where $\delta = \gamma-1$, we show that a choice of self-similar variable that exploits this compressive behavior in the limit that $\gamma \rightarrow 1$ naturally leads to a series expansion of the post-shock fluid density, pressure, and velocity in the small quantity $\delta$. We demonstrate that the leading-order (in $\delta$) solutions, which are increasingly accurate as $\gamma \rightarrow 1$, can be written in simple, closed forms when the fluid is still approximated to be in the energy-conserving regime (i.e., the Sedov-Taylor limit), and that the density declines exponentially rapidly with distance behind the shock. We also analyze the solutions for the bubble surrounding a stellar or galactic wind that interacts with its surroundings, and derive expressions for the location of the contact discontinuity that separates the shocked ambient gas from the shocked wind. We discuss the implications of our findings in the context of the dynamical stability of nearly isothermal shocks. 
\end{abstract}




\section{Introduction}
The injection of energy into a medium results in the formation of a shockwave: a discontinuity in the fluid properties that propagates outward from the explosion site. When the total energy of the explosion is conserved and the energy injection occurs impulsively, the flow is adiabatic\footnote{By this we mean that the only changes to the entropy of the post-shock fluid arise from the expansion of the gas.} and the Sedov-Taylor blastwave describes the temporal and spatial evolution of the post-shock fluid velocity, density, and pressure. This solution to the fluid equations was independently found by \citet{taylor50} and \citet{sedov59}, is self-similar in that it depends only on the relative position of a fluid element behind the shock front (modulo additional, multiplicative temporal scalings), and holds provided that the density profile of the ambient material into which the shock advances is a power-law in spherical radius  {(see also Section III of \citealt{ostriker88})}.

A distinct self-similar solution describing the propagation of a shock -- and the variation of the post-shock fluid variables -- is obtained when the energy injection into the fluid occurs mechanically (i.e., between two massive fluids) over a finite timescale. When the energy injection rate is constant, as would be the case for a constant luminosity wind that emanates from a stellar surface or a galaxy, the interaction between the wind supplying the energy and the ambient gas forms a ``bubble.'' This bubble consists of a forward shock that advances into the ambient gas, a reverse shock that travels back through the wind, and a contact discontinuity that separates the two shocked fluids (e.g., \citealt{castor75, weaver77, fielding18}); a self-similar solution describes the fluid properties between the forward shock and the contact discontinuity.

When the shock generated from a supernova explosion or an expanding wind is young, the large temperatures and densities at the shock front imply that the effective adiabatic index $\gamma$ of the multispecies fluid satisfies $4/3\lesssim \gamma \lesssim 5/3$. As the shock further expands and cools, however, the high-density regions near the shock front start to radiate efficiently, which reduces $\gamma$ below the value of an ideal monatomic gas. On the other hand, the low-density material in the interior of the explosion remains hot, and the expanding blastwave is characterized by a thin, nearly isothermal shell that is in contact with a hot, pressurized interior (e.g., \citealt{mckee77, shull79, blondin98}). 

Once the shock becomes radiative it is no longer strictly correct to treat the fluid as adiabatic. However, it is reasonable to assume that the gas exhibits an increased degree of compressibility prior to radiating a significant amount of its total energy, during which time it is still approximately correct to use the self-similar scalings (derived under strict adiabaticity) for the relationship between the shock position and velocity but with an appropriately reduced adiabatic index. At the very least, the self-similar solutions with a nearly isothermal equation of state describe the qualitative behavior we expect at this stage -- a nearly evacuated, extremely hot interior in contact with a thin shell that contains the majority of the mass (e.g., Figure \ref{fig:compsgam} below). Understanding the structure of these shocks therefore yields insight into the more general behavior of radiative, thin shells formed from explosions and stellar and galactic winds.

In this Letter, we argue that the self-similar variable typically used to derive self-similar solutions does not exploit the physical behavior of the blastwave when $\gamma \simeq 1$. In Section \ref{sec:scalings} we outline some basic physical considerations that suggest that the post-shock gas is contained in a shell of width $\Delta R/R \simeq (\gamma-1) \equiv \delta$, and we describe the adjusted self-similar variable that exploits the thinness of the shell when the adiabatic index of the gas nears unity. In Section \ref{sec:solutions} we provide the leading-order solutions for the fluid variables that result from an expansion of the fluid equations in the small quantity $\delta$; among our results is that the post-shock density declines exponentially rapidly behind the shock front when the fluid is in the Sedov-Taylor phase. We also apply our analysis to a wind-driven bubble and investigate the properties of the shell of material between the forward shock and contact discontinuity, and we find excellent agreement between our prediction for the location of the contact discontinuity that drives the shock to those in the literature. We offer concluding remarks in Section \ref{sec:conclusions}.

\section{Scalings and Basic Considerations}
\label{sec:scalings}
The fluid behind a shockwave obeys the equations of hydrodynamics, which in spherical coordinates are

\begin{equation}
\frac{\partial \rho}{\partial t}+\frac{1}{r^2}\frac{\partial}{\partial r}\left[r^2\rho v\right] = 0, \label{cont}
\end{equation}
\begin{equation}
\frac{\partial v}{\partial t}+v\frac{\partial v}{\partial r}+\frac{1}{\rho}\frac{\partial p}{\partial r} = 0, \label{rmom}
\end{equation}
\begin{equation}
\frac{\partial s}{\partial t}+v\frac{\partial s}{\partial r} = 0, \label{ent}
\end{equation}
where $r$ is the spherical radius, $\rho$ is the mass density, $v$ is the radial velocity, $p$ is the pressure, and $s = p/\rho^{\gamma}$ is the specific entropy with $\gamma$ the adiabatic index. We also ignored the gravitational potential energy, which assumes the fluid velocity is much greater than the freefall speed.\footnote{There are scenarios in which self-similar solutions may be found when gravitational effects are included; see \citet{coughlin18}}

To derive self-similar solutions to this set of equations, one usually makes the coordinate transformation

\begin{equation}
r \rightarrow \xi = \frac{r}{R(t)}, \label{xidef}
\end{equation}
where $R(t)$ is the time-dependent position of the shock. At the shock front the functions satisfy the jump conditions, which in the adiabatic regime (i.e., when radiative losses to not significantly modify energy conservation) give

\begin{equation}
v(R) = \frac{2}{\gamma+1}V, \,\,\, p(R) = \frac{2}{\gamma+1}\rho_{\rm a}(R)V^2, \,\,\, \rho(R) = \frac{\gamma+1}{\gamma-1}\rho_{\rm a}(R). \label{shockbcs}
\end{equation}
Here $V = dR/dt$ is the shock velocity, $\rho_{\rm a}(R)$ is the ambient density at the shock position, and we assumed that the shock velocity is much greater than the ambient sound speed. If the ambient density declines as a power-law in radius, so $\rho_{\rm a}(R) \propto R^{-n}$ with $n$ a constant, then we assume solutions to the fluid equations of the form

\begin{equation}
v = Vf(\xi), \,\,\, \rho = \rho_{\rm a}(R)g(\xi), \,\,\, p = \rho_{\rm a}(R)V^2h(\xi),
\end{equation}
where $f$, $g$, and $h$ are unknown functions. Inserting this ansatz into the fluid equations is self-consistent provided that $R\dot{V}/{V^2}$ is a constant. The three ordinary differential equations that result from the fluid equations can be integrated alongside the boundary conditions at the shock to yield the functions $f$, $g$, and $h$.

While this approach to solving the fluid equations is valid and self-consistent, it does not yield much insight into the behavior of the solutions themselves, and the choice of the self-similar variable $\xi$ as given in Equation \eqref{xidef} is not motivated by physical considerations. A more physical self-similar variable can be found by returning to the continuity equation \eqref{cont}, multiplying through by $r^2$, and integrating from some inner radius $R_0$ to just outside the shock $R(t)+\epsilon$. If we assume that the mass flux declines  {rapidly} as we move inward from the shock front, then the choice of the inner radius $R_0$ does not affect the solution; taking the limit as $\epsilon \rightarrow 0$ then gives

\begin{equation}
\frac{\partial}{\partial t}\int_{R_0}^{R(t)}r^2\rho dr = \rho_{\rm a}(R)R^2V \,\, \Rightarrow \,\, \int_{R_0}^{R(t)}r^2\rho dr = \frac{\rho_{\rm a}(R)R^3}{3-n}. \label{eq}
\end{equation}
The second equality results from the radial power-law dependence of the ambient density (see also Equation 4.11 of \citealt{ostriker88}). From the boundary conditions at the shock, however, the post-shock density is given by $\rho \simeq (\gamma+1)/(\gamma-1)\times \rho_{\rm a}(R)$; using this expression in the left-hand side shows that the mass behind the blastwave is contained in a shell of width $\Delta R$, where\footnote{This analysis is also valid when the shock is relativistic if one makes the replacement $\rho \rightarrow \rho\Gamma$ in the integral, where $\Gamma$ is the Lorentz factor of the fluid and $\rho$ is the comoving density; since the ultra-relativistic jump conditions give $\rho \propto \Gamma \rho_{\rm a}$, the thickness of the shell is $\Delta R/R \simeq 1/\Gamma^2$, in agreement with \citet{blandford76}.}

\begin{equation}
\Delta R \simeq \frac{1}{3-n}\frac{\gamma-1}{\gamma+1}R \equiv \frac{1}{3-n}\delta \times R,
\end{equation}
where we defined $\delta \equiv (\gamma-1)/(\gamma+1)$. We thus see that if $n$ is not too close to 3, then mass conservation alone tells us that the majority of the fluid is compressed into a shell of width $\sim \delta \times R$; this is also consistent with our neglect of the mass flux at $R_0$ in Equation \eqref{eq}. This observation suggests that a more physical self-similar variable, to be used in place of $\xi$, is

\begin{equation}
\chi = \frac{R-r}{\Delta R} = \frac{1}{\delta}\left(1-\frac{r}{R}\right) = \frac{1}{\delta}\left(1-\xi\right).
\end{equation}
Since $\delta$ is a constant  {(though see the discussion at the end of Section \ref{sec:conclusions})}, the Sedov-Taylor solutions in terms of $\chi$ are simply given by $f(\xi(\chi))$, etc. However, from the boundary conditions at the shock, we expect solutions to the fluid equations to be able to be written as a series in $\delta$, with the leading-order terms becoming increasingly accurate representations of the exact solution when $\delta$ is small and $\gamma\simeq 1$. In the next section we show not only that such a series expansion is possible, but that the leading-order solutions can be written in simple, closed form expressions when the energy contained in the shell is conserved. We also derive solutions for the shocked, ambient gas contained between the contact discontinuity and the forward shock in a wind-driven ``bubble.''

 \citet{ostriker88} extensively discuss the shell and thin-shell approximations to modeling blastwaves in more general contexts and note that the shocked material is highly compressed near the shock front when the gas becomes isothermal; see their Sections IIB, IVC, VA, and Appendix D. Our approach is more similar to the shell approximation, which maintains the finite thickness of the post-shock region, whereas the thin-shell model treats the shocked fluid as infinitely thin. However, here we account for the velocity structure within the post-shock fluid (see Equation \ref{fgheta} below; by contrast, the shell approximation treats the fluid velocity as exactly the post-shock velocity within the shell).

\section{Leading-order Solutions}
\label{sec:solutions}
\begin{figure*} 
   \centering
   \includegraphics[width=0.325\textwidth]{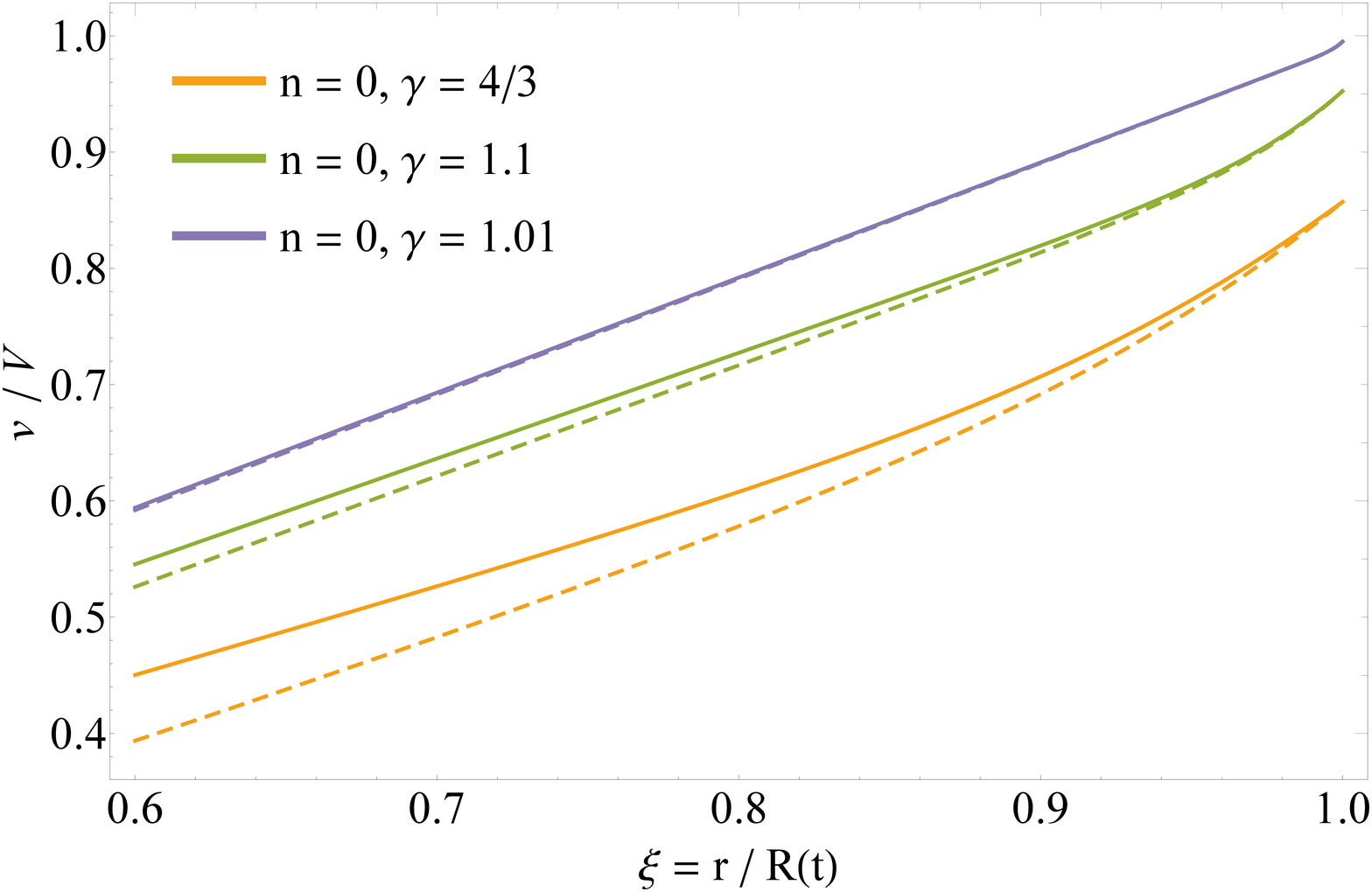} 
   \includegraphics[width=0.325\textwidth]{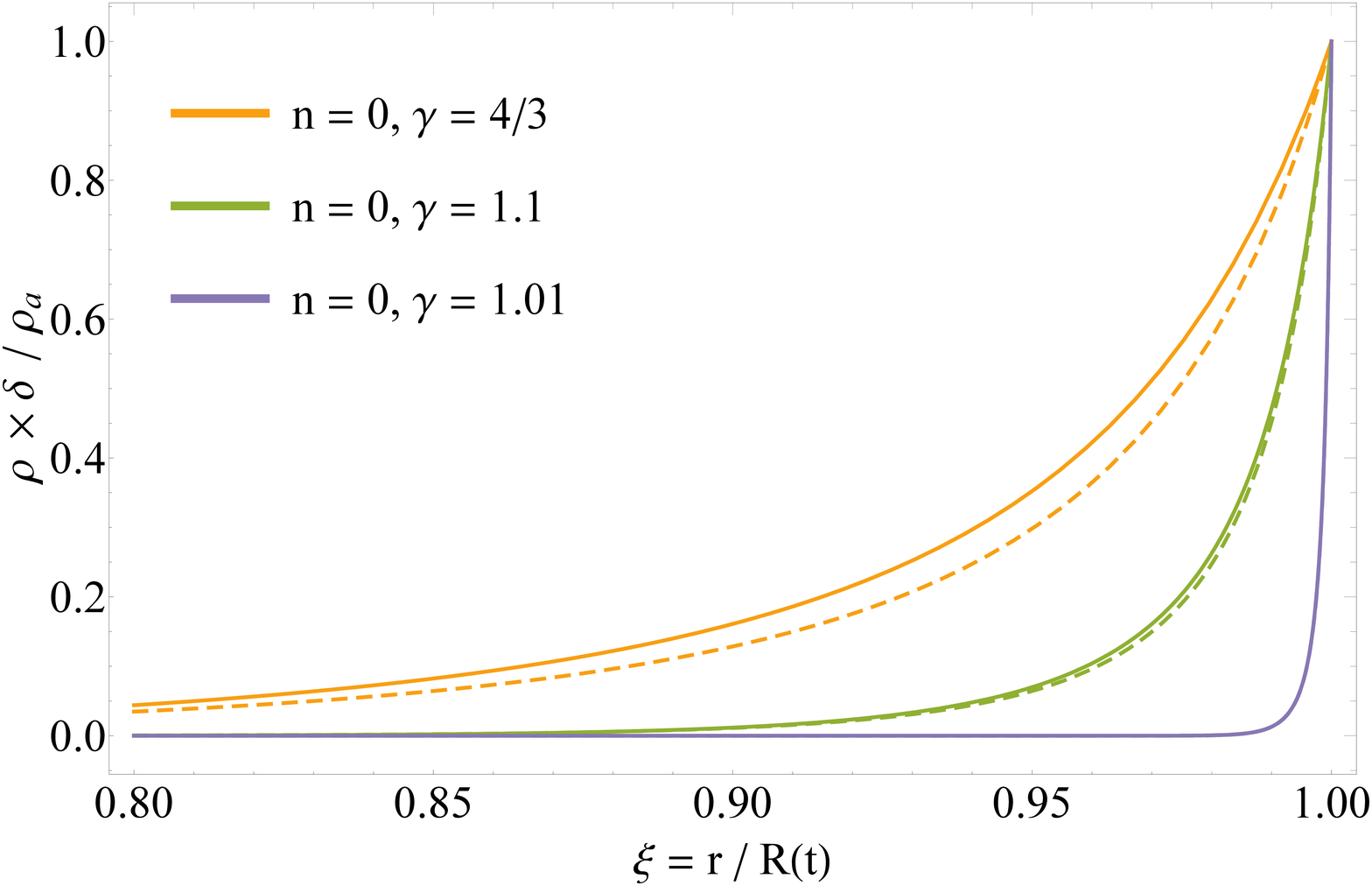} 
   \includegraphics[width=0.325\textwidth]{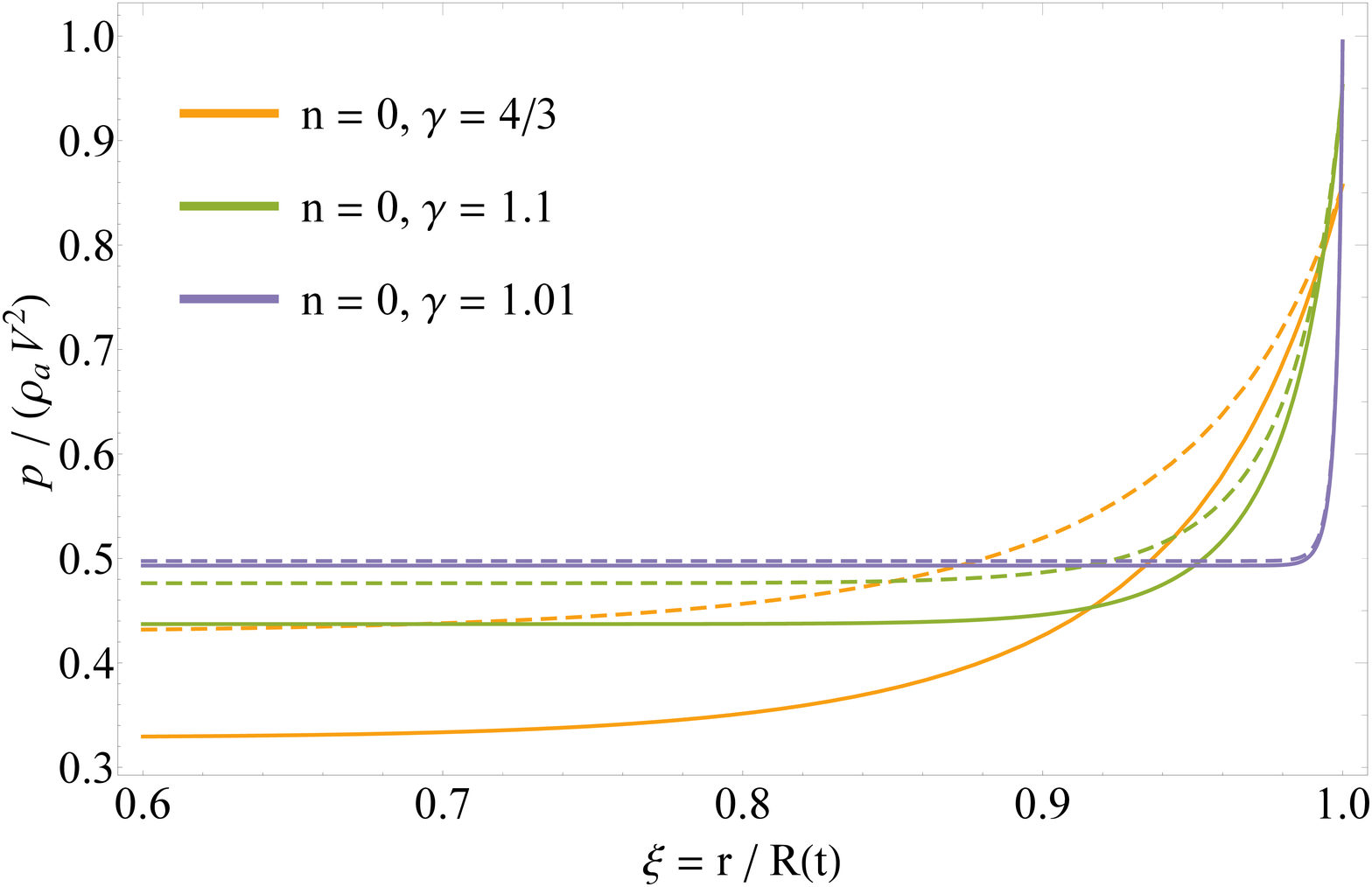} 
      \includegraphics[width=0.325\textwidth]{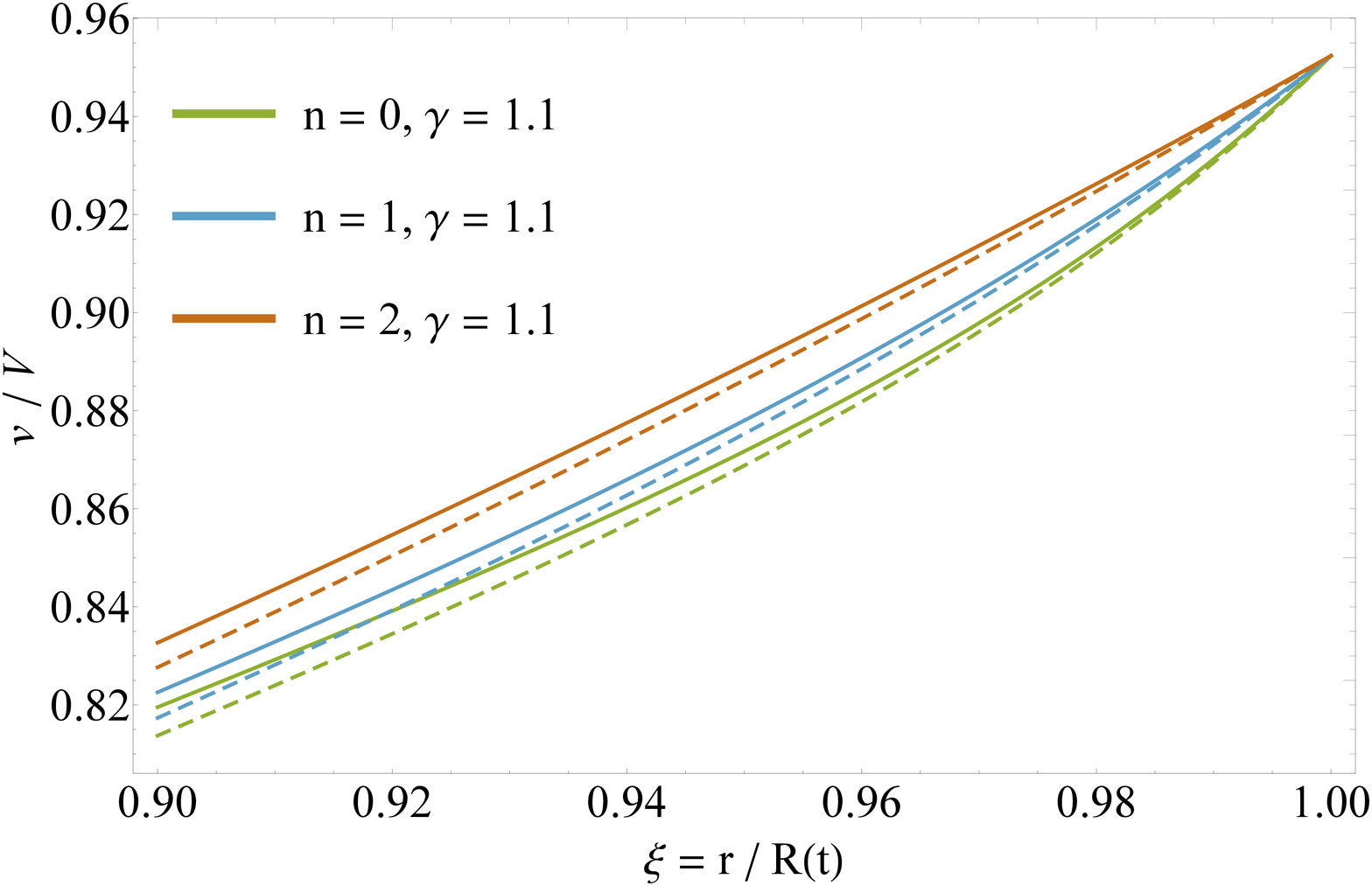} 
   \includegraphics[width=0.325\textwidth]{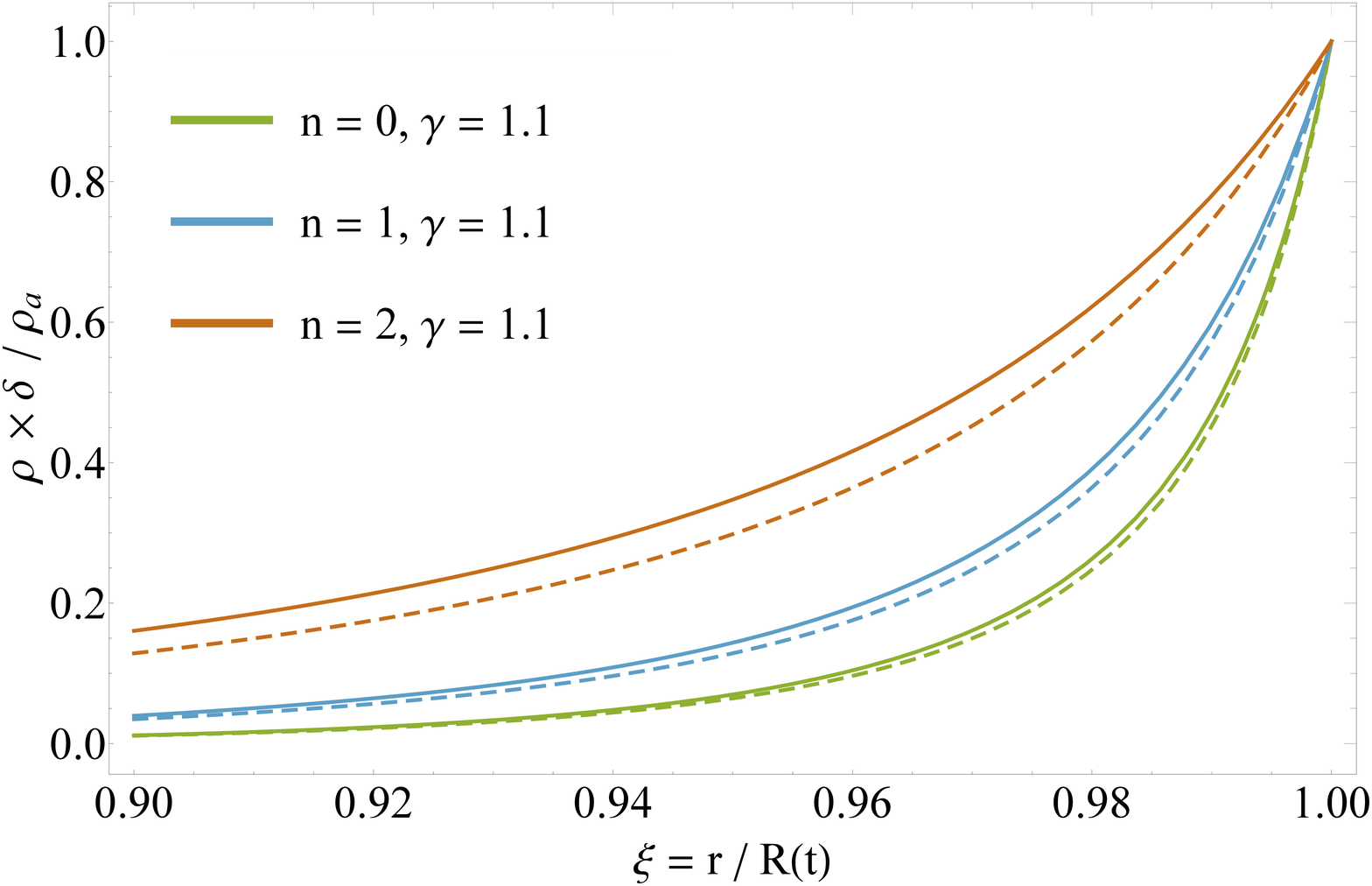} 
   \includegraphics[width=0.325\textwidth]{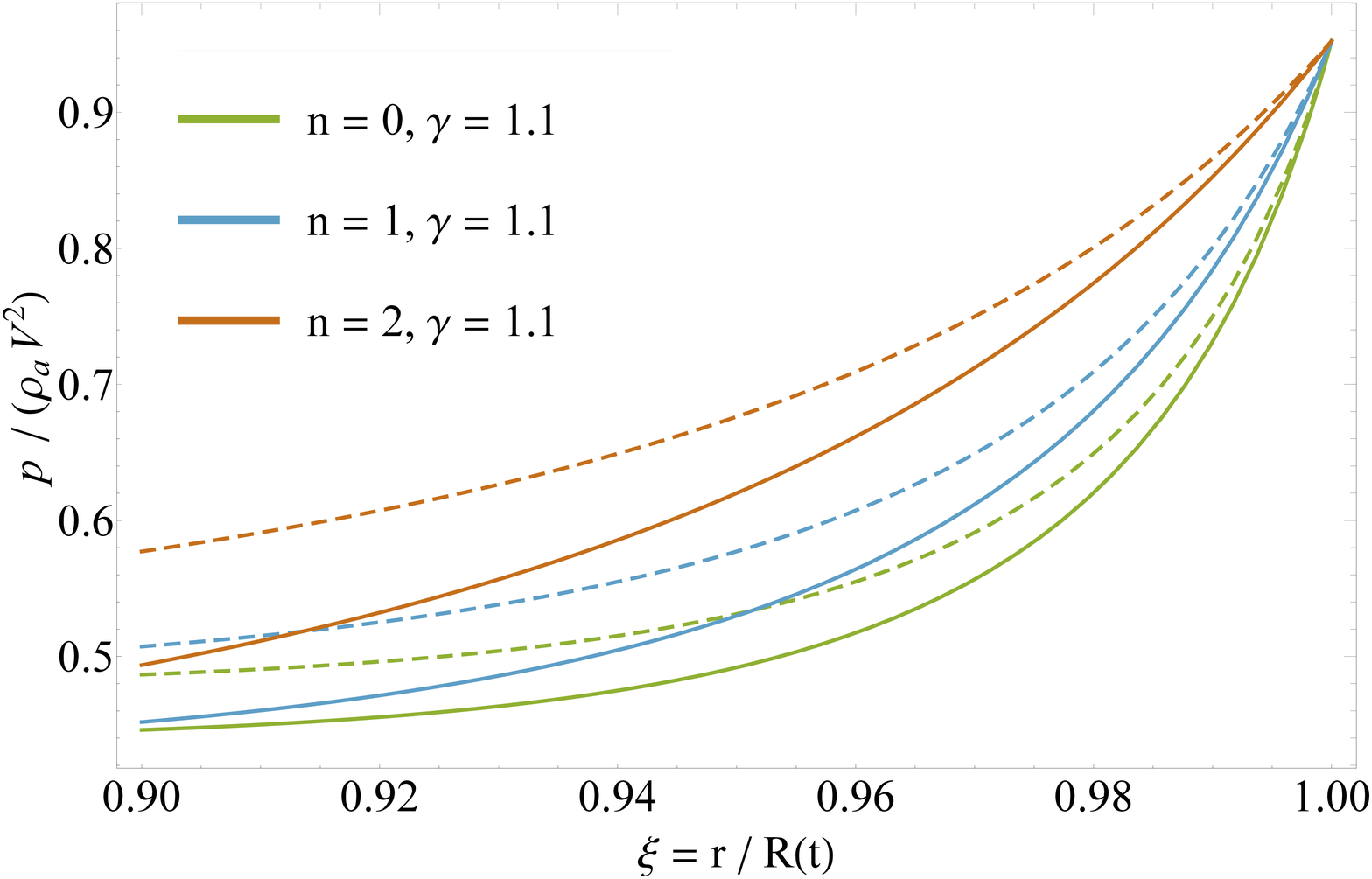} 
   \caption{A comparison of the exact solution for the Sedov-Taylor blastwave (solid lines) and the analytic, leading-order solution in the quantity $\delta = (\gamma-1)/(\gamma+1)$ given by Equation \eqref{sols} (dashed lines) for the adiabatic indices $\gamma$ and radial power-law indices $n$ shown in the legend; recall that the ambient density follows the power-law decline $\rho \propto R^{-n}$, where $R$ is the shock position. Solutions are plotted as functions of the standard Sedov-Taylor variable $\xi = r/R(t)$, which is just the spherical radius at a given time normalized by the shock radius. The left, middle, and right columns show the radial velocity normalized by the shock velocity, the density normalized by the ambient density (and reduced by the factor $\delta$), and the pressure normalized by the ram pressure, respectively. As $\gamma$ nears one, the analytic solutions provide better approximations for the structure of the shell.}
   \label{fig:compsgam}
\end{figure*}

Motivated by the preceding discussion and the boundary conditions at the shock front \eqref{shockbcs}, we write the fluid variables as

\begin{equation}
\begin{split}
&v = \left(1-\delta\right)V\left\{1+\delta f(\chi)\right\}, \\
&\rho = \frac{1}{\delta}R^{-n}g(\chi), \\ 
&p = \left(1-\delta\right)R^{-n}V^2h(\chi). \label{fgheta}
\end{split}
\end{equation}
These expressions should be interpreted as the leading-order terms in a series solution of the fluid variables in $\delta$ and $\chi$, and hence the exact solutions (e.g., the Sedov-Taylor solutions) will contain additional corrections that enter as higher powers of $\delta$; we can account for these terms by letting $f(\chi) \rightarrow f(\chi,\delta)$ and similarly for the other functions. However, when $\gamma$ is only marginally greater than one, we expect these leading-order expressions to be good approximations to the true solutions, and they represent the correct limiting forms as $\gamma \rightarrow 1$. The boundary conditions at the shock are now given by

\begin{figure*}
   \centering
   \includegraphics[width=0.325\textwidth]{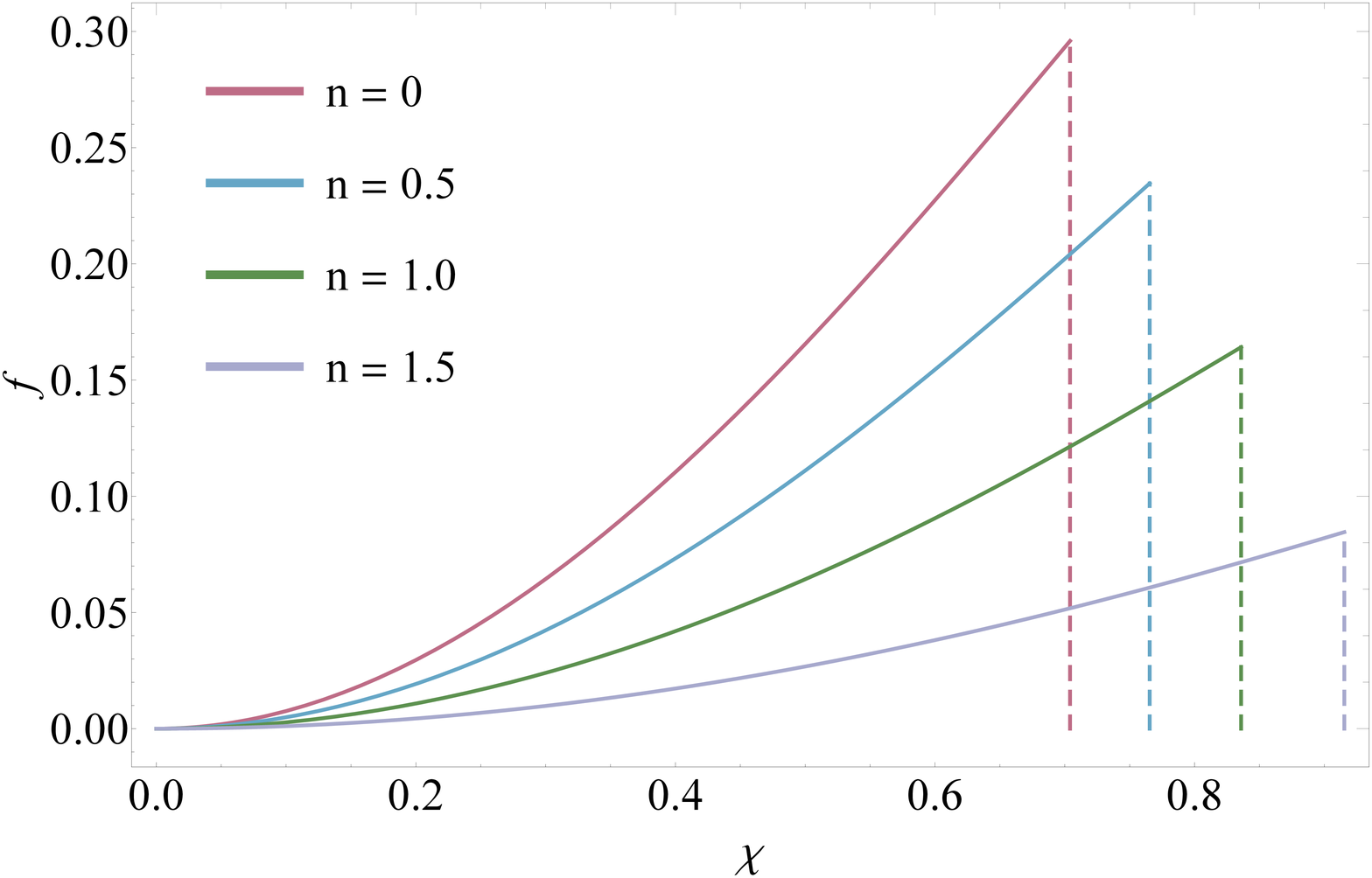} 
   \includegraphics[width=0.325\textwidth]{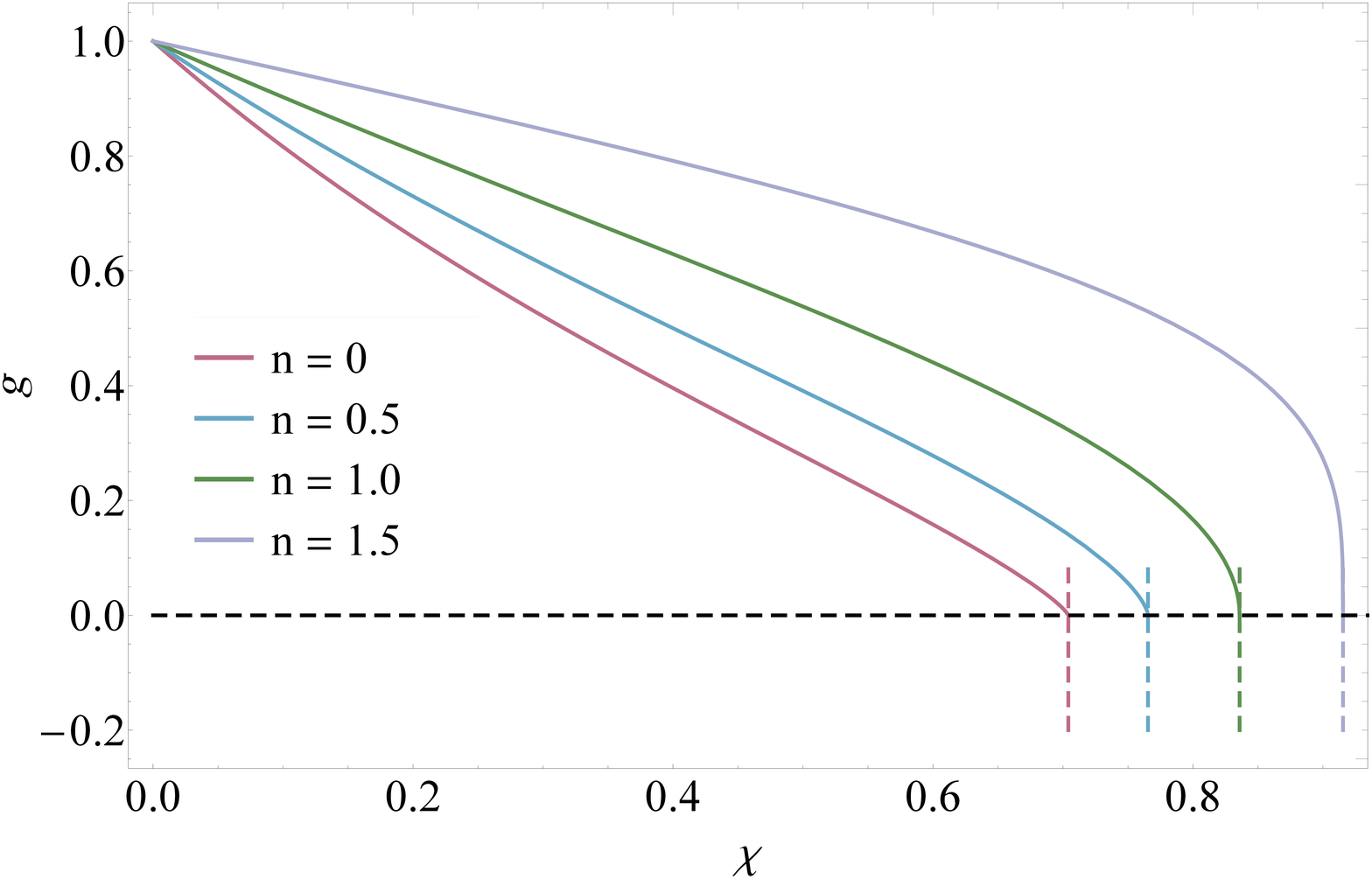} 
   \includegraphics[width=0.325\textwidth]{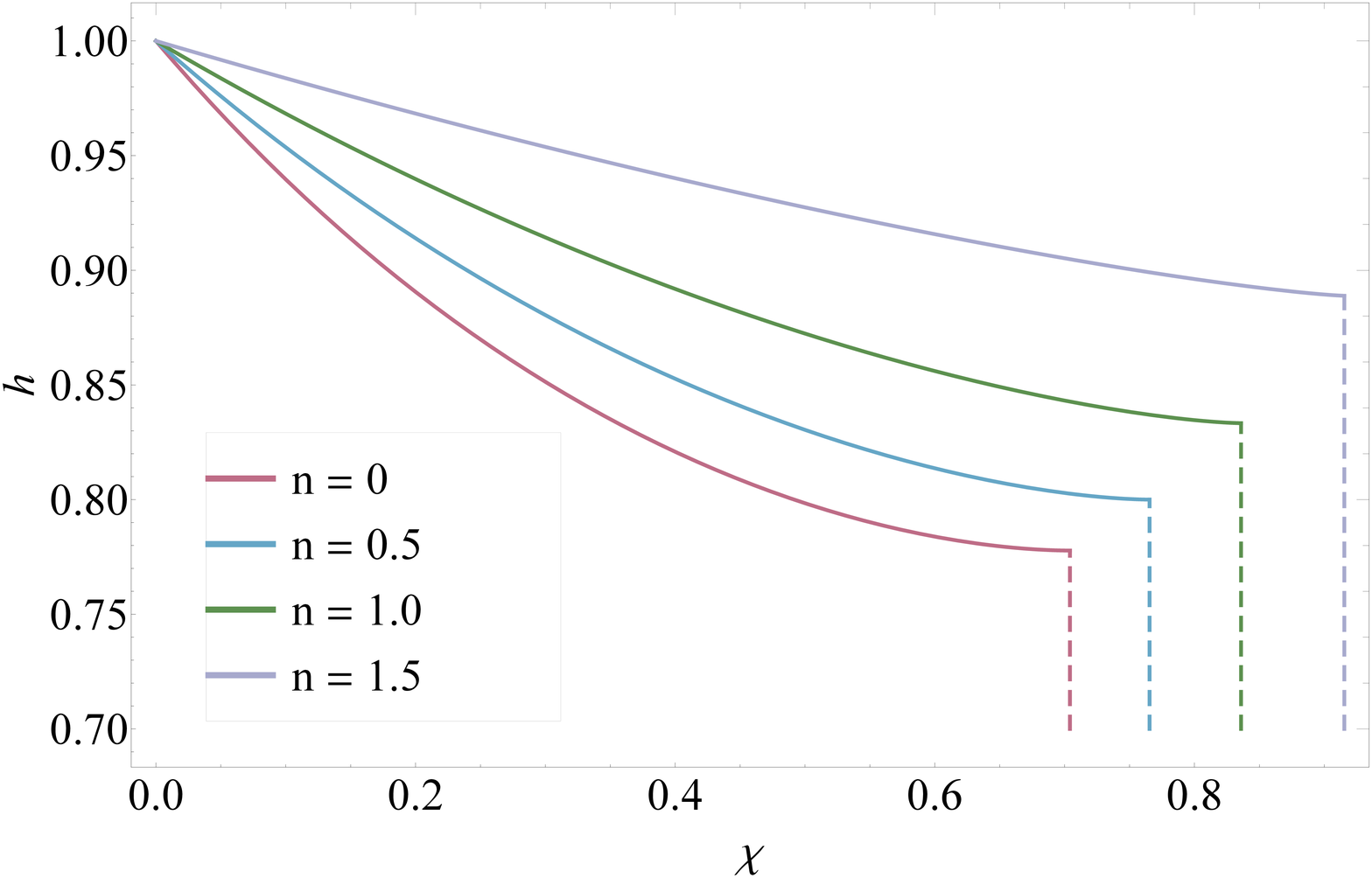} 
   \caption{The solutions for the normalized velocity (left), density (middle), and pressure (right) between the forward shock and contact discontinuity established by a wind-blown bubble. The different curves are for different density profiles of the ambient medium (i.e., the density of the ambient medium $\rho_{\rm a}$ scales with spherical radius $r$ as $\rho_{\rm a} \propto r^{-n}$). Here $\chi \propto 1-r/R$, and hence $\chi = 0$ corresponds to the location of the shock. As we advance into the shocked fluid ($\chi > 0$), the velocity increases slightly, the pressure remains nearly constant, and the density declines dramatically. The vertical lines show the location of the contact discontinuity that separates the shocked ambient gas from the shocked wind where the density equals zero. }
   \label{fig:bubble}
\end{figure*}

\begin{equation}
g(0) = h(0) = 1, \,\,\, f(0) = 0. \label{fgheta0}
\end{equation}
We can now insert the expressions for the velocity, density, and pressure in terms of the functions $f$, $g$, and $h$ (Equation \ref{fgheta}) into the fluid equations \eqref{cont} -- \eqref{ent}; doing so, keeping only leading-order terms in $\delta$, and making some algebraic manipulations gives

\begin{equation}
\left(3-n+\frac{du}{d\chi}\right)g = u\frac{dg}{d\chi}, \,\,\, \frac{R\dot{V}}{V^2} = \frac{1}{g}\frac{dh}{d\chi}, \,\,\, \frac{2R\dot{V}}{V^2} = -u\frac{d}{d\chi}\ln\left(\frac{h}{g}\right), \label{flred}
\end{equation}
where $u \equiv 1-f-\chi$. Here dots denote differentiation with respect to time. These equations are self-consistent provided that the quantity $R\dot{V}/V^2$ is a constant. We now analyze two cases where this constant is set by the total energy behind the blast being conserved (the Sedov-Taylor solution) and the shock being advanced by a contact discontinuity (wind-driven bubble solution).

\subsection{Energy-conserving blastwave}
\label{sec:sedov}
Since the total energy behind the shock scales as $V^2R^{3-n}$, imposing that this quantity be conserved gives $R\dot{V}/V^2 = (n-3)/2$. Using this expression for $R\dot{V}/V^2$ in Equation \eqref{flred}, the solutions are:

\begin{equation}
\begin{split}
& u = 2-e^{-\frac{3-n}{2}\chi} \quad \Leftrightarrow \quad  f (\chi)= -1-\chi+e^{-\frac{3-n}{2}\chi}, \\
&g(\chi) = \frac{2}{3-n}\frac{1}{u^2}\frac{du}{d\chi} = \frac{e^{-\frac{3-n}{2}\chi}}{\left(2-e^{-\frac{3-n}{2}\chi}\right)^2}, \\ 
&h(\chi) = \frac{1}{u} = \frac{1}{2-e^{-\frac{3-n}{2}\chi}}. \label{sols}
\end{split}
\end{equation}
The solid lines in Figure \ref{fig:compsgam} illustrate the Sedov-Taylor solutions for the post-shock fluid variables -- the normalized velocity (left column), density (middle column), and pressure (right column) -- which were numerically calculated from Equations (53) -- (55) of \citet{coughlin19}, while the dashed lines give the analytic solutions from Equation \eqref{sols}. The various adiabatic indices and radial power-law indices of the ambient medium (recall that the ambient density declines with spherical radius as $\rho_{\rm a} \propto r^{-n}$) are shown in the legend. Here solutions are plotted in terms of the usual Sedov-Taylor variable $\xi = r /R$ to facilitate the interpretation of the solutions. We see that the leading-order solutions provide good fits to the exact, numerically integrated solution that uses the variable $\xi = r/R$. 

These solutions demonstrate that as the adiabatic index nears unity, the limiting behavior of the density is to decline \emph{exponentially rapidly} behind the shock front, with an $e$-folding, dimensionless position of $\chi_{\rm e} = 2/(3-n)$; alternatively, in terms of the usual self-similar variable, the $e$-folding position is

\begin{equation}
\frac{r_{\rm e}}{R} \equiv \xi_{\rm e} = 1-\frac{\gamma-1}{\gamma+1}\frac{2}{3-n} \simeq 1-\frac{\gamma-1}{3-n},
\end{equation}
where in the last line we let $\gamma+1 \simeq 2$. We also see that the sound speed increases exponentially rapidly and the pressure gradient declines exponentially rapidly. 

 {We can also determine the Lagrangian evolution of the fluid elements behind the shock, which is governed by the relation $\partial r/\partial t = v$; using the solution for $f(\chi)$ in this relation yields}

\begin{equation}
\chi(\tau) = \frac{2}{3-n}\ln\left(\frac{1+e^{\left(3-n\right)\tau}}{2}\right), \label{chisol}
\end{equation}
 {where $\tau = \ln(R/R_0)$, $R_0$ being the initial position of the shock, and we imposed that the fluid element originates at the shock (i.e., $\chi(\tau = 0) = 0$). This result demonstrates that the appropriate time-like variable that describes the evolution of fluid elements within the post-shock flow is the total shock position $R$, not the width of the shell $\delta \times R$. In particular, for a constant-density medium ($n = 0$) a fluid element traverses the width of the shell $\delta \times R$, such that $\chi = 1$, after the shock expands by a factor of $\simeq 2$ (the exact factor can be found by setting $\chi = 1$ in Equation \ref{chisol} and rearranging). Physically this occurs because the fluid velocity in the comoving frame of the shock is reduced by a factor of $\delta$, and hence the shock expands by a factor of order unity by the time a fluid element traverses a distance $\delta \times R$.}

\subsection{Wind-driven bubble}
The choice of self-similar variable that leads to the boundary conditions at the shock (Equation \ref{fgheta0}) and the reduced set of fluid equations \eqref{flred} are valid when $R\dot{V}/V^2$ is a constant. If the total energy is conserved and imparted impulsively, $R\dot{V}/V^2 = (n-3)/2$, and the previous subsection shows that the solutions for the post-shock fluid variables can be written in simple, closed forms. 

Another possibility is that the energy reservoir driving the shock is in the form of a wind -- stellar or galactic -- that creates a ``bubble,'' which consists of a reverse shock, a forward shock, and a contact discontinuity. With a constant energy injection rate, the energy contained between the forward shock and the contact discontinuity increases linearly with time, which yields (since the energy is proportional to $V^2R^{3-n}$) $R\dot{V}/V^2 = (n-2)/3$. At late times the gas will become radiative and reduce the adiabatic index to $\gamma \simeq 1$, but energy losses will not dramatically offset the energy injection provided by the advancing contact discontinuity (and therefore change the value of $R\dot{V}/V^2$; e.g., \citealt{ryu88}). With $R\dot{V}/V^2 = (n-2)/3$, Equation \eqref{flred} will accurately describe the structure of the shell between the contact discontinuity and forward shock when $\gamma \simeq 1$.

Figure \ref{fig:bubble} shows the solutions for the dimensionless velocity $f$ (left panel), density $g$ (middle panel), and pressure $h$ (right panel) of the gas between the forward shock and the contact discontinuity. Here the solutions are plotted as functions of the self-similar variable $\chi \propto 1-r/R$, and hence the position of the shock coincides with $\chi = 0$ while $\chi > 0$ correspond to radii within the shocked shell. The different curves are appropriate to the radial power-law indices of the ambient medium, $n$, shown in the legend. We see that as we move inward from the shock front, the velocity remains nearly unchanged -- recall that the velocity is $\propto 1+\delta f \simeq 1$ -- while the pressure decreases slightly. The density declines from the shock and reaches zero at the contact discontinuity (vertical, dashed lines). 

Denoting the dimensionless position of the contact discontinuity by $\chi_{\rm c}$, these solutions demonstrate that the radial position of the contact discontinuity, $r_{\rm c}$, is

\begin{equation}
r_{\rm c} = \left(1-\frac{\gamma-1}{\gamma+1} \chi_{\rm c}\right)R \,\,\, \Rightarrow \,\,\, \frac{r_{\rm c}}{R} \equiv \xi_{\rm c} = 1-\frac{\gamma-1}{\gamma+1}\chi_{\rm c}, \label{rceq}
\end{equation}
$\xi_{\rm c}$ being the location of the contact discontinuity in terms of the usual self-similar variable. For $n = 0$, 0.5, 1.0, and 1.5, we have, respectively, $\chi_{\rm c} \simeq 0.704$, 0.765, 0.836, and 0.915. For a constant-density ambient medium ($n = 0$) and $\gamma = 1.05$, 1.1., 1.2, and 1.3, this result predicts $\xi_{\rm c} \simeq 0.983$, 0.966, 0.936, and 0.908, respectively. These values can be compared to the exact values given in Table 1 of \citet{ryu88}: for $\gamma = 1.05$ the two are identical (to the third decimal place), while for $\gamma = 1.3$ the difference is at the 1\% level (indeed, even for $\gamma = 5/3$ the prediction given by Equation \eqref{rceq} is discrepant from that in \citealt{ryu88} by only $\sim 3\%$). 

\section{Summary and Conclusions}
\label{sec:conclusions}
We argued that the standard self-similar variable $\xi = r/R(t)$ used to derive self-similar solutions to the fluid equations (e.g., the Sedov-Taylor solution), where $r$ is the spherical radius and $R(t)$ is the time-dependent position of an advancing shock, should be replaced by $\chi = (1-\xi)/\delta$, where $\delta = (\gamma-1)/(\gamma+1) \simeq (\gamma-1)/2$ when the shock is nearly isothermal ($\gamma \simeq 1$). Doing so manifestly exploits the physical structure of the post-shock fluid in this regime, being that all of the mass is compressed into a thin shell of width $\Delta R/ R \simeq (\gamma-1) \simeq \delta$. Using these arguments, we derived leading-order (in $\delta$), closed-form solutions for the Sedov-Taylor blastwave, which demonstrate that the post-shock density and pressure gradient decline exponentially rapidly behind the shock in the nearly isothermal limit. We also derived solutions when the shock is advanced by a radial wind (a forward shock, reverse shock, contact discontinuity ``bubble''), and found excellent agreement between our estimates of the location of the contact discontinuity and those that result from the exact solution. As demonstrated by Equations \eqref{flred}, the existence of these solutions depends only on the constancy of the combination of shock parameters $R\dot{V}/V^2$, where $V$ and $\dot{V}$ are the shock velocity and acceleration, respectively, and are thus valid for \emph{any} shockwave that satisfies this constraint (e.g., the shell generated during the interaction between supernova ejecta and a surrounding medium; \citealt{chevalier82}). 

When the shock enters this nearly isothermal phase, a dynamic instability can set in that destroys the self-similar nature of the shock \citep{vishniac83, ryu87, ryu88, sanz11}. Our approach to describing the self-similar solutions offers a distinct methodology for understanding the modes and the timescales over which such instabilities act. In particular, we note that only the leading-order terms in the series expansion of the fluid variables, Equation \eqref{fgheta}, need to be self-similar. When deriving the higher-order terms in $\delta$, we can therefore let $f(\chi) \rightarrow f(\chi)+\delta f_1(\chi,\tau)$ and similarly for the other variables, where $f_1$ is the next-order correction to the velocity in $\delta$ and $\tau = \ln R$  {(see the discussion at the end of Section \ref{sec:sedov})}. This transformation leads to linearized and separable equations in $\tau$ and $\chi$, from which we can derive the ``eigenvalues'' $\sigma$ that govern the oscillations of the fluid; perturbations then vary as $\sim e^{\sigma\tau}$. One such eigenvalue will be zero, as the exact solution (i.e., the Sedov-Taylor or wind-bubble solution) is time-independent and contains higher-order terms in $\delta$. The non-zero values give the non-trivial, fundamental modes of the shell, and -- purely from physical arguments -- they will vary as power-laws in the shock position $R$. 

Our analysis also shows that there is a smallness parameter $\delta \simeq \gamma-1$ that is the natural quantity about which to perturb the solutions. As the gas becomes increasingly isothermal this parameter becomes smaller, and hence any perturbations within, for example, the density profile of the ambient medium that modify the self-similar nature of the post-shock fluid variables must be smaller than or of the order $\delta$. If the perturbations are substantially larger than $\delta$, then these ``corrections'' to the self-similar nature of the flow become comparable to the self-similar, and unperturbed, solutions themselves. We therefore expect that fluctuations in the density of the ambient medium will more readily destroy the self-similar nature of the blastwave as the gas becomes increasingly isothermal, which agrees with detailed analyses \citep{ryu87}.  

 {When the adiabatic index of the gas is a constant, the solutions derived here are re-representations of the usual self-similar solutions (i.e., the Sedov-Taylor blastwave or the wind-driven bubble) written in terms of the canonical self-similar variable $\xi = r/R(t)$, and they demonstrate the appropriate limiting behavior of these solutions when $\gamma \rightarrow 1$. However, physically we expect the adiabatic index to be time-dependent, starting near $5/3$ and transitioning to a value near unity as the shock becomes radiative, and a model that would account for this effect would let $\delta \rightarrow \delta (t)$. Interestingly, we can show\footnote{ {Note that one must return to the general gas energy equation and include the explicit time dependence of the adiabatic index in this case; there is therefore an additional term that enters into the entropy equation \eqref{ent} that accounts for this variation.}} that if $\delta(t)$ behaves as a power-law in the shock position, so $\dot{\delta}/\delta = -k V/R$ (assuming that $\delta$ becomes smaller with time, so $k > 0$), and we make the same change of variables with $r\rightarrow \chi = (1-r/R)/\delta(t)$, then the leading-order (in $\delta$) fluid equations are \emph{precisely the same as Equation \eqref{flred}}, but with $u \rightarrow u+k\chi = 1-f-\chi+k\chi$. Thus, the self-similar functions for the pressure and density, $h$ and $g$, are precisely the same in this case, while the velocity declines less rapidly with distance behind the shock with increasing $k$ (more rapid cooling). However, these solutions are not simply redefinitions of the usual self-similar solutions, as the self-similar variable includes additional time dependence. }

\section*{Acknowledgements}
 {I thank the anonymous referee for a constructive report. I also thank my in-laws, Robert and Tina Berger, for providing me space to work while writing this paper.} Support for this work was provided by NASA through the Hubble Fellowship Program, grant \#HST-HF2-51433.001-A awarded by the Space Telescope Science Institute, which is operated by the Association of Universities for Research in Astronomy, Incorporated, under NASA contract NAS5-26555.

\bibliographystyle{mnras}
\bibliography{refs}

\bsp	
\label{lastpage}
\end{document}